\newcommand{\be}{\begin{equation}}
\newcommand{\ee}{\end{equation}}
\newcommand{\bea}{\begin{eqnarray}}
\newcommand{\eea}{\end{eqnarray}}
\newcommand{\ba}{\begin{array}}
\newcommand{\ea}{\end{array}}

\newcommand{\al}{\alpha}
\newcommand{\ga}{\gamma}
\newcommand{\Ga}{\Gamma}

\newcommand{\ka}{\kappa}
\newcommand{\de}{\delta}

\newcommand{\la}{\lambda}

\newcommand{\om}{\omega}
\newcommand{\Om}{\Omega}
\newcommand{\ze}{\zeta}
\newcommand{\De}{\Delta}
\newcommand{\La}{\Lambda}

\newcommand{\tha}{\theta}

\newcommand{\Sp}{\mathrm{Sp}}
\newcommand{\SU}{\mathrm{SU}}
\newcommand{\SO}{\mathrm{SO}}
\newcommand{\rL}{\mathrm{L}}

\newcommand{\R}{\mathrm{I}\kern -2.5pt \mathrm{R}}
\newcommand{\Z}{\mathsf{Z}\kern -5pt \mathsf{Z}}
\newcommand{\C}{\mathsf{I}\kern -5pt \mathrm{C}}

\newcommand{\D}{{\rm d}}
\newcommand{\pa}{\partial}
\newcommand{\rar}{\rightarrow}

\newcommand{\non}{\nonumber}
\newcommand{\we}{\!\wedge\!}

\newcommand{\cL}{\mathcal{L}}
\newcommand{\cF}{\mathcal{F}}

\newcommand{\cN}{\mathcal{N}}

\newcommand{\half}{\mbox{$\frac{1}{2}$}}
\newcommand{\bN}{\bar{N}}

\newcommand{\ts}{\textstyle}

\documentclass[12pt]{article}

\usepackage{amsfonts}
\usepackage{graphicx}

\newdimen\tableauside\tableauside=1.0ex
\newdimen\tableaurule\tableaurule=0.4pt
\newdimen\tableaustep
\def\phantomhrule#1{\hbox{\vbox to0pt{\hrule height\tableaurule
width#1\vss}}}
\def\phantomvrule#1{\vbox{\hbox to0pt{\vrule width\tableaurule
height#1\hss}}}
\def\sqr{\vbox{%
  \phantomhrule\tableaustep

\hbox{\phantomvrule\tableaustep\kern\tableaustep\phantomvrule\tableaustep}%
  \hbox{\vbox{\phantomhrule\tableauside}\kern-\tableaurule}}}
\def\squares#1{\hbox{\count0=#1\noindent\loop\sqr
  \advance\count0 by-1 \ifnum\count0>0\repeat}}
\def\tableau#1{\vcenter{\offinterlineskip
  \tableaustep=\tableauside\advance\tableaustep by-\tableaurule
  \kern\normallineskip\hbox
    {\kern\normallineskip\vbox
      {\gettableau#1 0 }%
     \kern\normallineskip\kern\tableaurule}%
  \kern\normallineskip\kern\tableaurule}}
\def\gettableau#1 {\ifnum#1=0\let\next=\null\else
  \squares{#1}\let\next=\gettableau\fi\next}

\newcommand{\Yfund}{\tableau{1}}

\begin{document}
 
\begin{flushright}
 BRX-TH-503 \\ {\tt hep-th/0206071}
\end{flushright}
\vspace{1mm}
\begin{center}{\bf\Large\sf An orientifold 
of $\mathit{adS}_5{\times}T^{11}$  with D7-branes, the  \\[-5pt] associated {\large $\al'^2$}-corrections and their role 
in the \\[3pt] dual {\large  $\cN=1$} {\large $\Sp(2N{+}2M){\times}\Sp(2N)$} 
gauge theory }
\end{center}
\vskip 3mm
\begin{center} 
Howard J. Schnitzer\footnote{Research 
supported in part by the DOE under grant
DE--FG02--92ER40706.},
and Niclas Wyllard\footnote{
Research supported by the DOE under grant DE--FG02--92ER40706.\\
{\tt \phantom{aaa} schnitzer,wyllard@brandeis.edu}\\}

\vspace*{0.3in}

{\em Martin Fisher School of Physics\\
Brandeis University, Waltham, MA 02454}

\end{center}
 
\vskip 5mm
 
\begin{abstract}
We study the $\cN=1$ $\Sp(2N{+}2M){\times}\Sp(2N)$ gauge theory on a stack of $N$ physical and $M$ fractional D3-branes in the background of an orientifolded conifold. 
The gravity dual is a type IIB orientifold   of $\mathit{adS}_5{\times}T^{11}$ (with certain background fluxes turned on) containing an O7-plane and 8 D7-branes. 
 In the conformal case ($M=0$), we argue that the $\al'^2$-terms localized on the 8 D7-branes and the O7-plane should give vanishing contributions to the supergravity equations of motion for the bulk fields. 
In the cascading case ($M\neq 0$), we argue that the $\al'^2$-terms give rise to corrections which in the dual  $\Sp(2N{+}M){\times}\Sp(2N)$ gauge theory can be interpreted  as corrections to the anomalous dimensions of the matter fields.
\end{abstract}

\setcounter{equation}{0}
\section{Introduction}
The duality between string theory on
$\mathit{adS}_5{\times} T^{11}$ and an $\cN=1$ $\SU(N){\times}\SU(N)$
gauge theory \cite{Klebanov:1998} (see \cite{Klebanov:2000c} for a review)
is one of the most interesting examples of a gauge/gravity dual pair
because the field theory has the minimal amount of supersymmetry. 
Furthermore, since $T^{11}$ is not locally $S^5$ the duality is not a simple extension of the basic \cite{Maldacena:1998} $\mathit{adS}_5{\times}S^5 \leftrightarrow \cN=4 \; \SU(N)$ correspondence.  
A more recent motivation to study this model comes from the fact that it is possible to break the conformal invariance by adding fractional branes to the defining D3-brane configuration, thereby changing the gauge group to $\SU(N{+}M){\times}\SU(N)$. In the gravity dual these fractional branes are replaced by certain fluxes.  
The modification of the model due to the addition of the fractional branes leads to several interesting features as demonstrated in \cite{Klebanov:2000b} (following earlier work in \cite{Klebanov:1999,Klebanov:2000a}, see also \cite{Herzog:2001,Herzog:2002}; other models within the same universality class have also recently attracted attention, see e.g.~\cite{Maldacena:2000}). 

Recently, the $\al'^3R^4$ corrections to the $\mathit{adS}_5{\times}T^{11}$ supergravity background  were studied  and it was shown that these corrections do not change the background when $M=0$ \cite{Frolov:2001}. 
This result supports the adS/CFT correspondence and leads to the conjecture that the $\mathit{adS}_5\!\times\! T^{11}$ supergravity solution (with a non-zero constant $\cF_5$ flux on $T^{11}$) should extend without modifications to an exact string theory background \cite{Frolov:2001}.
When $M\neq 0$ the $\al'^3R^4$ terms give rise to non-vanishing corrections to the background. It was argued in ref.~\cite{Frolov:2001} that the modification has an interpretation in the dual field theory as a correction to the anomalous dimension of the matter fields.

In this paper we study the $\cN=1$ $\Sp(2N{+}2M){\times}\Sp(2N)$ gauge theory on a stack of $N$ physical and $M$ fractional D3-branes in the background of an orientifolded conifold. This model has the following matter content: $2(\Yfund,\Yfund)\oplus4(1,\Yfund)\oplus 4(\Yfund,1)$, i.e. two bifundamentals, and 4 fundamentals for each of the two factors of the gauge group. The gravity dual is a type IIB orientifold of $\mathit{adS}_5{\times}T^{11}$ containing an O7-plane and 8 D7-branes \cite{Naculich:2001,Naculich:2002}. 
The fact that the orientifolded model contains D7-branes leads to several novel features not present in the unorientifolded model. One such feature is that both the coincident  D7-branes as well as the O7-plane have $\al'$ corrections localized on their world volume. 
The main theme of this paper is the study of the leading order corrections, which as we will see later occur at order $\al'^2$.  These corrections thus appear at a lower order in the $\al'$ expansion compared to the bulk $\al'^3$ terms. 
Since the charge of the D-brane is proportional to the inverse of the string coupling constant \cite{Polchinski:1995} these corrections also appear at a different order in the string coupling constant expansion. 
We argue that in the $M=0$ case (when the field theory is a superconformal theory) the $\al'^2$ terms give vanishing corrections to the equations of motion for the bulk fields. 
Our results are suggestive, but because of the lack of knowledge of the corrections to the equations of motion involving the Ramond-Ramond five-form gauge field $F_5$ we are unable to completely prove the result. 
The vanishing of the corrections would provide evidence that the orientifolded background with eight D7-branes is an exact string theory background and would also support the adS/CFT correspondence between string theory on  $\mathit{adS}_5{\times}T^{11}/\Z_2$ and the dual $\cN=1$ $\Sp(2N){\times}\Sp(2N)$ gauge theory. 
 A determination of the unknown $F_5$ terms would require the calculation of at least open string six- and eight-point functions; from the result one could then extract the $F_5$-dependent terms at order $\al'^2$. Such calculations are expected to be very involved and will not be attempted in this paper. 

In the cascading $M\neq 0$ case the $\al'^2$ terms are expected to lead to a modification of the background. We argue that in the field theory dual these corrections can be interpreted as corrections to the anomalous dimensions of the 
matter fields (as was the case for the bulk $\al'^3$ corrections \cite{Frolov:2001}). 
Apart from the lack of knowledge of the $F_5$ terms our analysis is further complicated because the background is less symmetric compared to the unorientifolded case and also because there are two species of matter fields: bifundamentals and fundamentals. Nevertheless, with suitable assumptions some progress can be made. 

In the next section we briefly discuss two possible orientifolded versions of the basic $\SU(N{+}M){\times}\SU(N)$ model and discuss the dual gravity backgrounds. One of the two models contains an O7-plane and 8 $D7$-branes. In section \ref{Corr1} we analyze the corrections to the equations of motion for the bulk fields induced by the $\al'^2$ terms localized on the O7 and the 8 $D7$'s in the conformal case, i.e.~when $M=0$. A similar analysis is carried out in section \ref{Corr2} for the case $M\neq 0$. In the appendix we discuss some string scattering amplitudes relevant to our analysis.

\setcounter{equation}{0}
\section{Orientifolds of $\mathit{adS}_5{\times} T^{11}$ and their field theory duals} \label{Sori}

The duality \cite{Klebanov:1998,Morrison:1998} between the $\cN=1$ $\SU(N){\times}\SU(N)$ conformal gauge theory with matter content $2(\Yfund,\overline{\Yfund})\oplus2(\overline{\Yfund},\Yfund)$, and string theory on $\mathit{adS}_5{\times}T^{11}$ can be extended to orientifolded versions thereof. 
The orientifolding can be chosen in such a way that it does not break any supersymmetry. There are two natural orientifolding operations, leading to models where the $\cN=1$ superconformal field theory duals have the following gauge groups and matter content
\be \label{models}
\ba{lllll} 
&\mathrm{(i)}& \Sp(2N){\times} \Sp(2N)\,,\qquad &\mathrm{with} \qquad &2(\Yfund,\Yfund)\oplus 4(\Yfund,1) \oplus 4(1,\Yfund) \,, \non \\
&\mathrm{(ii)}& \SO(2N{+}2){\times} \Sp(2N)\,,\qquad & \mathrm{with}\qquad &2(\Yfund,\Yfund) \,.
\ea
\ee
For later purposes it will be useful to recall some properties of the conifold and its relation to $T^{11}$. The conifold \cite{Candelas:1990} can be defined  by the equation $z_1^2 + z_2^2 + z_3^2 +z_4^2=0$ which describes a cone inside $\C^4$ (via a linear change of basis this can also be written as $xy -w z =0$). 
The base of the cone is  \cite{Candelas:1990} $T^{11} \equiv \frac{\SU(2){\times}\SU(2)}{\mathrm{U}(1)}$ and is obtained by intersecting the above space with $|z_1|^2 + |z_2|^2 + |z_3|^2 + |z_4|^2=1$. In a particular parameterization the metric of $T^{11}$ is given by \cite{Candelas:1990}
\be \label{T11met}
\D s^2 = \frac{1}{9}(\D \psi + \cos\tha_1\D \phi_1 + \cos \tha_2 \D \phi_2)^2 + \frac{1}{6}(\D \tha_1^2 + \sin^2\tha_1 \D\phi_1^2) + \frac{1}{6}(\D \tha_2^2 + \sin^2\tha_2 \D\phi_2^2)\,.
\ee
The dual string theory background corresponding to the models in (\ref{models}) is $\mathit{adS}_5{\times}T^{11}/\Z_2$, with a non-trivial self-dual five-form $F_5 = \cF_5 + *\cF_5$ reflecting the non-trivial $\cF_5$ flux on $T^{11}$. Here $\Z_2$ is the orientifold projection.
The action of the orientifold in the model whose field theory dual is $\Sp(2N){\times}\Sp(2N)$ was derived in \cite{Naculich:2001} (see also \cite{Naculich:2002}) to be $z\leftrightarrow w$ in terms of the conifold variables. In terms of the $z_i$'s it is $(z_1,z_2,z_3,z_4) \rar (z_1,z_2,z_3,-z_4)$. 
This result induces the following action on the coordinates of $T^{11}$\cite{Naculich:2001}: $\tha_1 \leftrightarrow \tha_2$ and $\phi_1 \leftrightarrow \phi_2$. The three-dimensional fixed point set of this action was called $X_3$ in \cite{Naculich:2001}. The model thus contains an O7 plane (and for consistency also 8 D7-branes coincident with the O7-plane). The world volume of the coincident O7 and D7's is $\mathit{adS}_5{\times} X_3$. 
 The metric on $X_3$ is
\be \label{X3}
\D s^2 = \frac{1}{9}(\D \psi + 2\cos \tha \D \phi)^2 + \frac{2}{6}(\D \tha^2 + \sin^2\tha \D\phi^2)\,.
\ee 
For the $\SO{\times} \Sp$ model the orientifold action can similarly be shown to be $x,y\rar -x,-y$ and $z \leftrightarrow w$, or in terms of the $z_i$'s: $(z_1,z_2,z_3,z_4) \rar (-z_1,-z_2,z_3,-z_4)$. This result was recently obtained in \cite{Imai:2001} (another orientifold operation was suggested in \cite{Ahn:2001}). For this case, the orientifold group does not have a fixed point inside $T^{11}$ so this model has no O-planes or D-branes. For other discussions of various other possible orientifolds of the conifold see e.g. \cite{Uranga:1998}.

The above orientifolded models (\ref{models}) can be extended by adding fractional branes in complete analogy with the unorientifolded case; the addition of such branes breaks the superconformal invariance and changes the gauge groups to $\Sp(2N{+}2M){\times}\Sp(2N)$ and $\SO(2N{+}2M{+}2){\times}\Sp(2N)$, respectively. Various aspects of the resulting models were recently discussed from the point of view of cascading gauge theories \cite{Imai:2001,Naculich:2002}

\setcounter{equation}{0}
\section{Stringy corrections I: the conformal case ($M=0$)} \label{Corr1}

In this section we discuss only the conformal field theories (i.e.~the $M=0$ case).  
The $\mathit{adS}_5\!\times\! T^{11}$ supergravity background (with a non-zero constant $\cF_5$ flux on $T^{11}$) is believed to extend without modifications to an exact string theory background \cite{Frolov:2001}. If true, this implies that all $\al'$ corrections to the leading supergravity equations of motion should vanish. This conjecture has not been proved, but the close relation to the conifold (which is an exact string tree-level background \cite{Witten:1993}) makes it plausible. 
String theory corrections would also lead to conflicts with the adS/CFT correspondence. For instance, there can be no corrections (modulo field redefinitions) to the equation of motion for the dilaton, since the dilaton has to remain a constant modulus in order not to conflict with the dual field theory being conformal. Similarly, the radius of $\mathit{adS}_5{\times}T^{11}$, which is related to $g_{\mathrm{S}}N$, must remain a free parameter in the full string theory \cite{Frolov:2001}. 
In \cite{Frolov:2001} it was checked that the leading corrections to the type IIB supergravity theory --- the famous $\al'^3 R^4$ terms \cite{Gross:1986} ---  do not give rise to any corrections to Einstein's equation or to the equation of motion for the dilaton\footnote{This had been checked earlier for the $\mathit{adS}_5\times S^5$ background \cite{Banks:1998}, for which the proof is much simpler. }. 
More precisely, it was shown that there exists a particular choice of field-redefinition scheme in which this is true. 
It was also pointed out that there could in principle be additional corrections to the equations of motion at order $\alpha'^3$ involving $F_5$. The result in \cite{Frolov:2001} implies that such terms, if present,  have to vanish separately (assuming that the background {\it is} exact).

In the previous section we discussed two orientifolds of the $\mathit{adS}_5 {\times} T^{11}$ background and their field theory duals. 
For these theories there are also $\al'^3 R^4$ corrections to the bulk supergravity action which are of the same form as in the unorientifolded theory, the only difference arising from the fact that the orientifold restricts the ranges of some of the variables. 
Consequently it follows from the result in \cite{Frolov:2001} that they give no corrections to the equations of motion. 

For the $\mathit{adS}_5 {\times} T^{11}/\Z_2$ orientifold corresponding to the $\Sp(2N){\times}\Sp(2N)$ theory there is, however, another source of $\al'$ corrections. Recall that this model contains 8 D7-branes and one O7-plane.  There are terms of order $\al'^2$ localized on the $\mathit{adS}_5{\times} X_3$ world-volume of these branes/plane, which  appear at a lower order in the $\al'$ expansion than the bulk $\al'^3$ terms. 
Since the charge of the D-brane is proportional to the inverse of the string coupling constant \cite{Polchinski:1995} these corrections also appear at a different order in the string-coupling constant expansion\footnote{There are also subleading $\al'^3$ corrections localized on the D/O7's, but since these are of a different order in $g_{\mathrm{S}}$ they can be treated separately and will not affect the above argument about the bulk $\al'^3$ terms.}. Our goal is to discuss what effects these $\al'^2$ corrections have on the equations of motion for the bulk fields. Before turning to this issue we will review some properties of embeddings, with particular emphasis on the case relevant to our discussion: the embedding of the world volume of the O7-plane and the D7-branes inside ten-dimensional space.

\subsection{Properties of embeddings}

For the discussion of how the world volume of the O7-plane and the D7-branes is embedded into ten-dimensional space the following notation will be used. Indices in the ten-dimensional (bulk) directions will be labeled by capital roman letters from the middle of the alphabet ($M,N,\ldots$); indices in the $\mathit{adS}_5$ directions will be labeled by lower case roman letters from the beginning of the alphabet ($a,b,\ldots$); indices in the $X_3$ directions will be labeled by greek letters ($\mu,\nu,\ldots$) and finally indices normal to $\mathit{adS}_5{\times}X_3$ will be labelled by $i,j,\ldots$ (note that these conventions differ from the ones used in \cite{Naculich:2001}).

The case relevant to our discussion is the embedding\footnote{We really mean the embedding of $\mathit{adS}_5{\times}X_3$ inside $\mathit{adS}_5{\times}T^{11}$, but since the embedding of the $\mathit{adS}_5$ part is trivial it will be suppressed in this section.} of $X_3$ inside $T^{11}$. This embedding is described by the vectors $\pa_{\mu}Y^M$ tangent to $X_3$ and the vectors $\xi_i^M$ normal to $X_3$. These two sets of vectors satisfy the relations:
\be \label{rels}
\xi^M_i G_{MN} \pa_{\mu}Y^N =0\,, \qquad \de_{ij} = \xi_i^M G_{MN} \xi_j^N\,, \qquad g_{\mu\nu}:= \pa_{\mu}Y^MG_{MN}\pa_{\nu}Y^N\,,
\ee
where $g_{\mu\nu}$ is the induced metric on $X_3$. The various types of indices are raised and lowered with $G_{MN}$, $g_{\mu\nu}$ and $\de_{ij}$. 
We will also use the tensors (defined on the world volume of the branes)
\be \label{LN}
L^{MN} := \pa_{\mu}Y^M g^{\mu\nu} \pa_{\nu}Y^N\,, \qquad N^{NM}:=\xi_i^N \de^{ij}\xi_j^M \,.
\ee 
It follows from (\ref{rels}) that these two symmetric tensors are orthogonal projection operators ($L_N{}^R L_R{}^M = L_N{}^M$, $N_N{}^R N_R{}^M = N_N{}^M$ and $L_N{}^R N_R{}^M = 0$). They are related to the restriction of the bulk metric to the world-volume via the relation $G^{MN} = L^{MN} + N^{NM}$. 
For the present case we can  choose (the coordinates of $T^{11}$ are labelled as $X^M = (\psi,\phi_1,\tha_2,\phi_2,\tha_2)$ and the coordinates on $X_3$ as $x^{\mu} = (\psi,\phi,\tha)$)
\be \label{dYs}
\pa_{\psi}Y^M = \left[ \ba{c} 1 \\ 0 \\ 0 \\ 0 \\ 0 \ea \right]\,;\qquad 
\pa_{\phi}Y^M = \left[ \ba{c} 0 \\ 1 \\ 0 \\ 1 \\ 0 \ea \right]\,;\qquad
\pa_{\tha}Y^M = \left[ \ba{c} 0 \\ 0 \\ 1 \\ 0 \\ 1 \ea \right]\,;
\ee
and 
\be
\xi_{1}^M = \frac{\sqrt{3}}{\sin{\tha}}\left[ \ba{c} 0 \\ 1 \\ 0 \\ -1 \\ 0 \ea \right]\,;\qquad 
\xi_{2}^M = \sqrt{3}\left[ \ba{c} 0 \\ 0 \\ 1 \\ 0 \\ -1 \ea \right]\,.
\ee
It is straightforward to check that the induced metric obtained using (\ref{dYs}) agrees with the expression given in (\ref{X3}) and that all the relations listed above are satisfied.

Bulk tensors can be projected to the normal and tangent directions by contracting with $\pa_{\mu}Y^M$ and $\xi_i^M$ and then restricting to the world volume; alternatively, one can use $L^{MN}$ and $N^{NM}$ and work only with tensors with bulk indices.  We use a notation where the occurrences of $\pa_{\mu}Y^M$ and $\xi_i^M$ are suppressed. The following definitions will be used below 
\be \label{Rs} \ba{lll}
 &R_{\rho\la\mu\nu} :=  \pa_{\rho}Y^P\pa_{\la}Y^Q \pa_{\mu}Y^M\pa_{\nu}Y^N R_{PQMN} \,,  &
 R_{\mu\nu} := \pa_{\mu}Y^Q  \pa_{\nu}Y^N  L^{PM} R_{PQMN}\,,  \non \\
& R_{ij\mu\nu} := \xi_i^P \xi_j^Q \pa_{\mu}Y^M\pa_{\nu}Y^N R_{PQMN}\,, &
 R_{ij} := \xi_i^Q  \xi_j^N  L^{PM} R_{PQMN}  \,.
\ea
\ee
An important tensor in the theory of embeddings is the second fundamental form (also known as the extrinsic curvature), defined as
\be \label{Om}
\Om_{\mu\nu}^M := \pa_{\mu}\pa_{\nu} Y^M - (\Ga_{\mathrm{T}})^{\rho}_{\mu\nu}\pa_{\rho}Y^M + \Ga^{M}_{NP}\pa_{\mu}Y^N\pa_{\nu}Y^P \,.
\ee
Here $(\Ga_{\mathrm{T}})^{\rho}_{\mu\nu}$ is the connection constructed out of the induced metric, $g_{\mu\nu}$, whereas $\Ga^{M}_{NP}$ is the connection constructed out of the bulk metric, $G_{MN}$.
Using the above formul\ae{} a straightforward calculation shows that $\Om^M_{\mu\nu}$ vanishes for the embedding of $X_3$ inside $T^{11}$ (thus the embedding is an example of what is referred to as a totally geodesic embedding). 

We should also mention that the curvatures of the tangent bundle, $(R_T)_{\rho\la\mu\nu}$, and the normal bundle, $(R_N)^{ij}{}_{\mu\nu}$, common in the literature, are related to certain pull-backs of the bulk Riemann tensor together with quadratic expressions in the second fundamental form:
\bea \label{RTRN}
&& (R_T)_{\rho\la\mu\nu} = \pa_{\rho}Y^P\pa_{\la}Y^Q \pa_{\mu}Y^M\pa_{\nu}Y^N R_{PQMN} + \xi_N^i \de_{ij} \xi_M^j[\Om^M_{\rho\mu}\Om^N_{\la\nu} - \Om^M_{\rho\nu}\Om^N_{\la\mu}] \,, \non \\
&& (R_N)_{ij\mu\nu} =  \xi_i^P \xi_j^Q \pa_{\mu}Y^M\pa_{\nu}Y^N R_{PQMN} + \xi_i^M g^{\rho\la} \xi_j^N [\Om_{M\rho\mu}\Om_{N\la\nu} - \Om_{N\rho\nu}\Om_{M\la\mu}] \,.
\eea
 For further details about embeddings relevant to our discussion, see e.g. \cite{Bachas:1999,Carter:1997}.

\subsection{The $\al'^2 R^2$ terms in the world-volume theories}

Because the charge of one O7-plane cancels that of 8 D7-branes and because there are no corrections at order $\al'$, the leading terms in the effective action for the combined O7-D7 system appear at order $\al'^2$. 
 We will now discuss these $\al'^2$ terms, which are localized on the O7-plane and the D7-branes. Since the only non-vanishing fields in the background we are discussing are $\Phi$, $G_{MN}$ and $F_{MNPQR}$, we are interested in the dependence of the effective action on these fields\footnote{In general the action for the D-branes also depends on the world-volume fields (i.e.~the gauge field and the transverse scalars; no such fields are present on the non-dynamical O7-plane). We will in what follows assume that these fields are zero. See section \ref{disc} for a discussion about this assumption.}. 
Because of symmetries, the other supergravity fields appear at least quadratically in the action. 
Since the zeroth-order expressions vanish, such terms will not affect the leading linearized corrections to the equations of motion. This last statement is actually not true for the Wess-Zumino term which is linear in the $C_p$'s; this fact will be taken into account in our later analysis. 
Unfortunately, the complete expression for the $\al'^2$ corrections is not known; in particular, the terms involving $F_5$ are not known. 
We will therefore concentrate on the corrections involving the Riemann tensor and the dilaton, which are known (some previously unknown terms will be determined in this paper). The effective action also depends \cite{Bachas:1999} on the second fundamental form (\ref{Om}). 
However, $\Om^M_{\mu\nu}$ appears at least quadratically so these terms will not affect our discussion since $\Om^M_{\mu\nu}$ is zero for the embedding we are discussing and we are only interested in the first order variation.

For the Wess-Zumino term of a D7-brane the relevant corrections involving the Riemann tensor take the form \cite{Bershadsky:1996} (ignoring the dependence on $\Om^I_{\mu\nu}$ for reasons just discussed)
\be
S^{\mathrm{D7}}_{\mathrm{WZ}} = \mu_7 ({\ts \frac{\pi^2}{6}})\al'^2 L^4 g_{S}^{-1} \int C_4 \we [ \mathsf{R}^{\rho}{}_{\la} \we \mathsf{R}^{\la}{}_{\rho} - \mathsf{R}^i{}_j \we \mathsf{R}^j{}_i ]\,,
\ee
where $\mathsf{R}^{\rho}{}_{\la} = \frac{1}{2!}R^{\rho}{}_{\la\mu\nu}\D x^{\mu}\we\D x^{\nu}$, $\mathsf{R}^i{}_j = \frac{1}{2!}R^{i}{}_{j\mu\nu}\D x^{\mu}\we\D x^{\nu}$, and we have rescaled $C_4\rar 4 g_{\mathrm{S}}^{-1}L^4C_4$, and $G_{MN}\rar L^2 G_{MN}$ to agree with the usual adS/CFT normalization (where $L$ is the $\mathit{adS}_5{\times}T^{11}/\Z_2$ length scale, $L^4 = {(27 \ka N)}/{(16 \pi^{5/2})}$). In the above formula $\mu_7$ is the charge of a single D7-brane, $\mu_7 = (2(2\pi)^7\al'^4)^{-1}$. 
The D-branes we discuss appear in an orientifolded theory and have {\it half} the charge of a D-brane in the type II theory (this is the reason for the factor of 2 discrepancy between the above equation and the expression usually found in the literature). 

The relation between the charge of an O7 orientifold plane and that of a D7-brane is $\mu^{\mathrm{O7}} = -8\mu^{\mathrm{D7}}$ (thus the charge of 8 D7's cancels that of one O7). The terms involving the Riemann tensor in the Wess-Zumino term for an O7-plane take the form \cite{Dasgupta:1998a}
\be
S^{\mathrm{O7}}_{\mathrm{WZ}} = (-8\mu_7)({\ts -\frac{\pi^2}{12}})\al'^2 L^4 g_{S}^{-1} \int C_4 \we [  \mathsf{R}^{\rho}{}_{\la} \we \mathsf{R}^{\la}{}_{\rho} - \mathsf{R}^i{}_j \we \mathsf{R}^j{}_i ]\,.
\ee
For the DBI part of the D7-brane action, the $R^2$ corrections were determined in \cite{Bachas:1999} (see also \cite{Fotopoulos:2001}) to be\footnote{The contributions involving the second fundamental form are ignored for the reason given above, the leading $\sqrt{-g}$ contribution is also ignored, as it will cancel against analogous terms coming from the O7-plane in the final expression.}
\be
S^{\mathrm{D7}}_{\mathrm{DBI}} = \mu_7 L^4 {\ts \frac{\al'^2}{2}(\frac{\pi^2}{6}) }\!\int \! \D^8x \sqrt{-g}\,e^{-\Phi}\Big[ R_{\rho\la\mu\nu}R^{\rho\la\mu\nu} - 2 R_{\mu\nu}R^{\mu\nu}  -\; R_{ij\mu\nu}R^{ij\mu\nu} + 2R_{ij}R^{ij}\Big] \,,
\ee
where the various tensors have been defined in (\ref{Rs}) and as above we have rescaled the metric as $G_{MN} \rar L^2 G_{MN}$.

The $\al'^2$ corrections to the DBI part of the O7-plane action have not, as far as we are aware, been discussed in the literature. We argue in the appendix that the expression for these terms is in line with expectations: 
\be \label{ODBI}
S^{\mathrm{O7}}_{\mathrm{DBI}} = (-8\mu_7)L^4{\ts \frac{\al'^2}{2}(-\frac{\pi^2}{12}) } \!\int\! \D^8x \sqrt{-g}\,e^{-\Phi}\Big[ R_{\rho\la\mu\nu}R^{\rho\la\mu\nu} {-} 2 R_{\mu\nu}R^{\mu\nu}  {-}\; R_{ij\mu\nu}R^{ij\mu\nu} {+} 2R_{ij}R^{ij}\Big] \,.
\ee
Adding the above terms, the total action for the combined system of 8D7's and one O7 becomes (in the string frame)
\bea  \label{R2tot}
S^7_{\mathrm{tot}} &=&  \mu_7 L^4 (\pi\al')^2 \Bigg\{ \int \! \D^8x \sqrt{-g}\,e^{-\Phi}\Big[ R_{\rho\la\mu\nu}R^{\rho\la\mu\nu} - 2 R_{\mu\nu}R^{\mu\nu}  -\; R_{ij\mu\nu}R^{ij\mu\nu} + 2R_{ij}R^{ij}\Big] \non \\ &&\qquad \quad +\; 2 g_{S}^{-1} \int C_4 \we [  \mathsf{R}^{\rho}{}_{\la} \we \mathsf{R}^{\la}{}_{\rho} - \mathsf{R}^i{}_j \we \mathsf{R}^j{}_i ] \Bigg\}\,.
\eea
Below we investigate how the presence of these terms affect the equations of motion for the bulk supergravity fields. There are also terms in the effective action of the schematic form $R D^2\Phi$ which contribute to the equation of motion for the dilaton; such terms are discussed below and in the appendix.

\subsection{The equations of motion}

In this section we will discuss the corrections to the equations of motion for the bulk supergravity fields arising from the presence of the terms (\ref{R2tot}). For the general discussion we restrict ourselves to the dependence of the action on the background metric.  In the case relevant to us, the action for the combined bulk+brane system takes the form (in the Einstein frame)
\be \label{b+b}
\int \D^{10}X\sqrt{-G}\, R + \int \D^{8}x \cL_{\mathrm{brane}} = \int\D^{10}X \sqrt{-G}\,R + \int \D^{10}X \de(\beta)\de(\ga) \cL_{\mathrm{brane}}\,,
\ee
where $\beta$ and $\ga$ label the transverse coordinates. In the background we are interested in, the following translations between the coordinates of $T^{11}$ and the  coordinates normal and tangential to $X_3$ are useful
\bea
&&\tha = (\tha_1+\tha_2)/2 \,, \qquad \phi = (\phi_1+\phi_2)/2\,, \qquad \psi = \psi \,,\non \\
&&\beta = (\tha_1-\tha_2)/2 \,, \qquad \ga = (\phi_1-\phi_2)/2\,.
\eea
The equations of motion are obtained by making a (eulerian) variation of the background metric: $\de G_{MN} = h_{MN}$. 
In the case when $\cL_{\mathrm{brane}}$ only depends on the metric and not on terms with derivatives, the brane gives rise to a simple $\de$-function source in Einstein's equation,
\be \label{eed}
\sqrt{G}[R_{MN} -\half G_{MN} R] = \de(\beta)\de(\ga) T_{MN} \,, \quad \mathrm{where} \; T_{MN} = \frac{\de \cL_{\mathrm{brane}}}{\de G^{MN}}\,.
\ee
 This means that the brane is treated as being infinitesimally thin. When the action also includes terms with derivatives acting on $G_{MN}$ (as is the case for the action we are interested in, cf.~(\ref{R2tot})), the situation is more involved. 
In the variation, a derivative acting on $\de G_{MN} = h_{MN}$ can be either in a direction tangential to $X_3$ or in a direction normal to it. If it is in a tangential direction then it can be moved by an integration by parts in the eight-dimensional brane action as usual. 
Hence, if the action only involves tangential derivatives then the formula (\ref{eed}) remains valid. 
On the other hand, if a derivative acting on $\de G_{MN}$ is in a  direction normal to $X_3$, it can still be moved by an integration by parts, but only in the {\it full ten-dimensional} action (i.e.~in the expression on the right-hand side of eq.~(\ref{b+b}), which includes the $\de$-functions). 
When such an integration by parts is performed, the derivative will hit the $\de$-functions. Thus, in general one gets not only contributions to the equations of motion involving $\de$-function source terms but in addition also source terms involving derivatives of $\de$-functions. 
The presence of such ``multipole'' source terms signals a violation of the approximation that the brane is infinitesimally thin (for a discussion of this point see \cite{Carter:1997}, section 3.5 and references therein and also \cite{Larsen:1997}). In general such corrections lead to complicated sources and the equation of motion may not have well-behaved solutions. 
Perhaps the best way to view such a situation is to say that when the solutions to the equation of motion become ill-defined, then the approximation of writing the action as a sum of two terms (bulk+brane) is no longer viable. However, as long as the solutions are well-behaved the approximation should still be 
viewed as being valid. (Note that solutions singular at the location of the brane need to be carefully regularized when pulled back to the brane world-volume, see \cite{Carter:2002} for a recent discussion.)

In the case of interest to us there will be at most two derivatives acting on $\de G_{MN} = h_{MN}$. Since one is to restrict the expression to the brane world-volume after acting with the derivatives, it is clear that it is sufficient to consider an expansion of $h_{MN}$ to second order in the transverse coordinates $\beta$ and $\ga$ (which are set to zero when restricting to the brane world volume): 
\be \label{deG}
h_{MN} = h^1_{MN}(\tha) + \beta h^2_{MN}(\tha) +  \ga h^3_{MN}(\tha) + \beta^2 h^4_{MN}(\tha) 
+ \beta \ga h^5_{MN}(\tha) + \ga^2 h^6_{MN}(\tha)\,.
\ee
As a technical simplification, we have indicated that it is sufficient to take the $h^i_{MN}$'s to depend on $\tha$.  
Since the  zeroth-order metric does not depend on $\psi$ and $\phi$, expressions involving derivatives with respect to these coordinates will give zero after an integration by parts. (The variation in general also depends on the radial coordinate of $\mathit{adS}_5$; we have suppressed this dependence in (\ref{deG}), but it will be needed for the general analysis below.)

In the variation the contributions from the different $h^i_{MN}$'s in (\ref{deG}) can be treated independently, since they give rise to sources with different $\de$-function structures. 
Because of the different possible source terms, there are (at order $\al'^2$) six times as many equations compared to the bulk case that have to vanish in order for there to be no corrections to the  equations of motion. The constraints are thus stronger than the ones arising from the bulk terms in \cite{Frolov:2001} and should lead to a more stringent test of the exactness of the background. At higher orders in the $\al'$ expansion one gets even further conditions since there are more derivatives present so additional ``multipole'' sources become relevant.

After these general remarks we are in a position to check whether the presence of the terms in (\ref{R2tot}) lead to corrections to the equations of motion. 
We will discuss each of the non-zero background fields in turn.
For the analysis it is important to keep in mind that one is allowed to make field redefinitions. 
Thus, to show that the $\al'^2$ terms in (\ref{R2tot}) do not give any corrections to the equations of motion, it is sufficient to find a field redefinition scheme in which the corrections vanish. 
In the bulk case there was a natural way to fix the field redefinition ambiguity: namely, to simply replace the Riemann tensor whenever it appeared with the Weyl tensor. 
In our case it is not a priori clear what field redefinition scheme is the correct one to use. 
In particular, replacing the bulk Riemann tensor with the Weyl tensor would make the relation (\ref{RTRN}) invalid (the pullback of the Weyl tensor can not be expressed in terms of $R_T$ in a simple way). 
Another ambiguity is which tensor one should contract with to form the ``Ricci'' tensors  $R_{\mu\nu}$ and $R_{ij}$ in (\ref{Rs}). Because of these problems we will in our analysis  add the most general combination of terms removable by field redefinitions to the action (\ref{R2tot}). 
Terms proportional to the lowest order equations of motion are removable by field redefinitions. For the metric the equation of motion is $R_{MN}=0$ and for the dilaton it is $D^M\pa_M\Phi=0$ (in a particular scheme \cite{Metsaev:1987}). These are actually not the complete expressions, but the additional terms will not be relevant to us. For instance there are $F_5^2$ corrections to the equation $R_{MN}=0$ but since the $F_5$ dependent $\al'^2$ terms are not known we will ignore the $F_5$-dependent corrections to the lowest order equations of motion as well. There are also dilaton corrections of the form $\pa_M \Phi \pa_N \Phi$, but as these are quadratic in $\pa_M\Phi$ they will not affect our analysis since the dilaton is constant to leading order.

\subsubsection{The equation of motion for $C_4$}
Since we are ignoring the unknown dependence of the $\al'^2$ terms on $F_5$, $C_4$ appears  only in the Wess-Zumino term. 
To get a correction to the equation of motion for $C_4$ from these $\al'^2$ terms, the $\mathsf{R}\wedge \mathsf{R}$ part in $C_4\wedge \mathsf{R}\wedge\mathsf{R}$ has to be non-vanishing  to leading (zeroth) order. For this to be the case $\mathsf{R}\wedge \mathsf{R}$ has to  have all its indices in the $adS_5$ directions. The reason for this is that because of the product structure of the zeroth order metric the leading order contribution to  $\mathsf{R}\wedge \mathsf{R}$ has to have all its indices either in the $adS_5$ directions or in the $X_3$ directions, but since it is a four-form it will not fit inside $X_3$. 
Thus, the only possible non-zero contribution at order $\al'^2$ to the equation of motion for $C_4$ comes from the variation with respect to $C_{abc\mu}$. 
However, all possible terms  vanish because for $\mathit{adS}_5$ one has: $R_{abcd} \propto G_{a[c}G_{d]b}$. It then follows that for any term that one writes down, the form indices will always be on the metric and there will always be a contraction either of the form $L^{c d} G_{c[a}G_{b]d}$ or of the form $N^{c d} G_{c[a}G_{b]d}$, both of which give zero. (Thus we get no information about which choice should be made to fix the field redefinition ambiguity.) 
We  conclude that there are no corrections to the equations of motion for $C_4$. This statement of course has to be revisited once the $\al'^2$ terms involving $F_5$ have been determined.

Apart from $C_4$ the Wess-Zumino term also has a linear dependence on the other $C_{p}$'s and on $B_2$. However, none of them are multiplied by quantities that are non-vanishing for the background under discussion. Thus there are no corrections to the equations of motion for the other $C_p$'s or $B_2$ either.

\subsubsection{The equation of motion for $\Phi$}
We next turn to the equation of motion for the dilaton.  
One contribution to the equation of motion comes from the variation of the $e^{-\Phi}$ factor appearing in the DBI part of the action ($\ref{R2tot})$. In addition to the overall $e^{-\Phi}$ factor, there could also be terms in the action involving derivatives of the dilaton. 
Such terms have not been determined in the literature. In the appendix we derive constraints on the $R \,D^2\Phi $ corrections to the DBI part of the action from the NSNS two-string scattering amplitude, cf. (\ref{PhiR}). When varying the $R D^2 \Phi$ terms one should expand $\Phi$ in ``modes'' as was done for the metric variation in (\ref{deG}).  We have checked that there is a particular field-redefinition scheme for which the corrections to the equation of motion coming from the $R D^2\Phi $ terms corresponding to all the six modes of $\Phi$ vanish.  The contribution from the mode independent of the normal coordinates should be added to the variation of the overall $e^{-\Phi}$ factor. However, there is no scheme in which this mode gives a non-vanishing contribution when the contributions from the other modes are required to vanish. 
 We have similarly checked that there are no corrections from the possible $R \, D^2 \Phi$ corrections to the Wess-Zumino term. 
To conclude, there is one constraint coming from the variation of the overall $e^{-\Phi}$ factor. This constraint can certainly be satisfied by fixing the field redefinition ambiguity of the $R^2$ terms suitably, but the resulting constraint on the coefficients has to be added to similar conditions arising from the variation of the metric.

\subsubsection{The equation of motion for $G_{MN}$} 
Finally we address the question of whether corrections to the equation of motion for the background metric are induced by the terms (\ref{R2tot}).  As we discussed above there are six different components of the variation (\ref{deG}); each of these has to vanish separately. The variation is most easily implemented when the action is written in terms of $L^{MN}$ and $N^{NM}$ (\ref{LN}) (rather than $\de_{\mu} Y^M$, $\xi_i^M$, $\de^{ij}$, and $g^{\mu\nu}$). One has
\be
\de L^{MN} = -L^{MP} \de G_{PQ} L^{QN}\,,\qquad \de N^{MN} = -G^{MP}\de G_{PQ}G^{QN} + L^{MP} \de G_{PQ} L^{QN}\,.
\ee
For the Wess-Zumino term, the only dependence on the metric is contained in the $\mathsf{R} \we \mathsf{R}$ terms. At zeroth order $C_4$ (choosing the natural gauge) only has non-vanishing components in either the $adS_5$ or the $T^{11}$ directions (but no mixed components). 
Therefore, the pullback of $C_4$ to the world volume has to lie in the adS directions (it will not fit inside $X_3$), which implies that the $\mathsf{R}\we \mathsf{R}$ terms have to have one index in an adS direction (and the others in the $X_3$ directions). 
At first sight one might think that this is not possible because of the product structure of the zeroth order metric, but the first order variation does not have to be zero since the adS index is a form index which is not contracted with the metric. 
There are two possibilities: (i) the adS index is on the metric, $G_{a M}$; or (ii) the adS index is on a partial derivative, $\pa_a$. 

For the first case, it is easy to see that the other index on the metric can not be a form index and that it has to be a $T^{11}$ index otherwise one gets zero because of the product  of the zeroth-order metric. 
Furthermore, the product structure of the zeroth order metric implies that to get something  non-vanishing the variation has to act on $G_{aM}$. Using this result together with the definition of the Riemann tensor one can check that the variation becomes ($\de G_{aM} = h_{aM}$)
\bea
(\mathrm{i}) \quad \de R_{PQaN} &\rar& \half G_{PS}\Big\{ -\pa_{N}(G^{SU}[\pa_Q h_{a U} - \pa_{U}h_{a Q}])
+ \Ga^T_{NQ}(G^{SU}[\pa_T h_{a U} - \pa_{U}h_{a T}]) \non \\ && \qquad \quad -\; \Ga^S_{NT}(G^{TU}[\pa_Q h_{a U} - \pa_{U}h_{a Q}]) \Big\}\,.
\eea
Expanding $h_{a M} = \de G_{aM}$ as in (\ref{deG}) one in general gets 30 conditions since the index $M$ in $h_{aM}$ can take 5 values and each $h_{aM}$ has six components, cf. (\ref{deG}). 
However, because of the symmetries of the $T^{11}$ there are smaller number of conditions, and we have checked that all of them are satisfied. 
In this calculation we also allowed for terms removable by field redefinitions. Such terms can be generated by replacing the factors of $R_{PQMN}$ by either $G_{P[M}R_{N]Q} - G_{Q[M}R_{N]P}$ or $R G_{P[M}G_{N]Q}$, where $R_{MN}$ is the bulk Riemann tensor, and $R$ is a scalar constructed out of it by contraction (recall that terms proportional to $R_{MN}$ can be removed by field redefinitions; strictly speaking, this is only true in the absence of $F_5$, but we ignore all $F_5$ dependence). 
Since such terms also give zero we get no clues which field redefinition scheme should be chosen.

For the second possibility (when the adS index is on a $\pa$) the variation will only give something non-vanishing when it acts on the terms which are acted on by the $\pa_a$. The reason for this is that the zeroth order metric on $T^{11}$ does not depend on the coordinates of $\mathit{adS}_5$, but the first order variation of the metric in general does. Since the $T^{11}$ metric does not depend on the adS coordinates, one can integrate by parts so that $\pa_a$ acts on  $C_4$. Thus the effect of this integration by parts is to turn $C_4$ into $F_5$. Using the definition of the Riemann tensor together with the above restriction on the variation one can show that  
\bea
(\mathrm{ii}) \quad \de R_{PQaN} &\rar& \pa_a\Big\{ G_{PS}\de \Ga^S_{NQ} + \half \Ga^T_{NQ}\de G_{TP} - \half G_{PS}\pa_N (G^{SU}\de G_{QU}) \non \\ && \quad -\; \half G^{TU}G_{PS} \Ga^S_{NT}\de G_{QU}\Big\}\,.
\eea
Because of the special background and our normalizations, $(F_5)_{abcde}$ is simply the volume form on $\mathit{adS}_5$. The contributions coming from the Wess-Zumino term obtained in this way are actually of the same form as terms coming from the DBI part of the action so they will have to be considered together. One thing to note is that the terms quadratic in the coordinates $\beta$, $\ga$ normal to $X_3$ in (\ref{deG}) will not contribute to the variation of the Wess-Zumino term. The reason for this is that at most one derivative acts on $\de G_{MN}$ since the other derivative is in the adS directions and has been integrated by parts. On the other hand such corrections are obtained from the variation of the DBI part of the action.

Finally, we will discuss the variation of the DBI part of the action. Because of the product structure of the zeroth order metric the variation has to lie either inside $\mathit{adS}_5$ or inside $T^{11}$ (there are no $\de G_{aM}$ contributions). The variation of the metric takes the form (\ref{deG}). In the case where the variation is in the $T^{11}$ directions the first three terms in (\ref{deG}) have counterparts coming from the Wess-Zumino term as discussed above, whereas the last three terms are only relevant for the DBI part. 
By analyzing all the above different equations with the help of the GRTensor package \cite{GrTensor}, we have found that there exists a choice of field-redefinition scheme for which the corrections involving the modes quadratic in the normal coordinates vanish. For the remaining modes (including the condition discussed above coming from the equation of motion for $\Phi$) there does {\em not} exists a field redefinition scheme\footnote{We have checked that this is still true even if one includes terms removable by field redefinitions of the schematic form $D^2R$.} in which there are no corrections to the equation of motion for the bulk metric coming from (\ref{R2tot}). However, there does exist an action giving rise to no corrections which is very similar to (\ref{R2tot}), but with slightly different numerical coefficients in front of the various terms (more precisely, two of the coefficients need to be different from the values in (\ref{R2tot}) to get a vanishing result).  The implications of this result will be discussed in the next section.

\subsection{Discussion of the result} \label{disc}
As discussed above there does not exist a field-redefinition scheme in which the entire variation with respect to the metric gives a vanishing result. 
Since we neglected the $F_5$-dependent terms, the fact that we did not find that all the corrections vanished is perhaps not too surprising especially considering that the restrictions imposed are stronger and the number of possible terms is smaller compared to the bulk analysis in \cite{Frolov:2001}. 
For instance, the number of terms removable by field redefinitions is smaller than the number of constraint equations coming from requiring the corrections to vanish. To complete the analysis one would need to take into account also the dependence of the action on $F_5$. In \cite{Frolov:2001} the corrections vanished even without the $F_5$ terms (hence the terms involving $F_5$ have to vanish separately), in our case the situation appears to be different. 

The equation of motion for the metric is expected to involve terms which depend on the self-dual RR field $F_5$. The various possible terms are (schematically) of the form $D F_5 D F_5$, $F_5^2 R$ and $F_5^4$. These terms have not been determined (the corrections to the DBI action of the first type can be constrained  from the scattering amplitudes in \cite{Gubser:1996,Garousi:1996}). To determine these terms from string scattering amplitudes one would have to calculate at least six- and eight-point open string scattering amplitudes. 
Needless to say that is a very involved calculation and we will not attempt it here. An important question is whether the additional $F_5$-dependent terms are such that they {\it can} make the corrections vanish. At first sight it seems that the $F_5$ terms could easily modify the above result since there are several possible tensor structures.  However, the number of independent contributions is actually much less than one might naively expect as can be seen e.g.~from the fact that because of the product structure of the zeroth order metric,  
the simple form of $F_5$ ($\propto \mathrm{vol}_{\mathit{adS_5}} + \mathrm{vol}_{T^{11}}$) together with the fact that the indices are contracted with either $L^{NM}$ or $N^{NM}$ (which are orthogonal projection operators) the various possible $F_5^4$ terms actually give rise to only three independent terms which can be chosen as $c_1 L^{ab}\de G_{ab} + c_2L^{MN}\de G_{MN} + c_3N^{MN}\de G_{MN}$, where the indices in the first term only run over the adS directions and the $c_i$'s are constants. 
Results such as this indicate that the $F_5$ terms can not significantly change the $R^2$ analysis. Thus, the fact that the corrections almost vanished using the $R^2$ terms only is a strong indication that when the $F_5$ terms are included one will find the extra contributions to the coefficients required for the vanishing of all corrections. This would be a very strong test of the exactness of the background. Finally, we should mention that we have assumed that the gauge field and the transverse scalars on the brane are zero. This may be incorrect, non-zero values may be required e.g.~for the solution to be supersymmetric (to analyze whether this is the case, the methods in \cite{Maldacena:2000b} might be useful). If non-zero world-volume fields are present they will only affect the modes of the variation of the metric independent of the coordinates normal to the brane.

\setcounter{equation}{0}
\section{Stringy corrections II: the cascading case ($M\neq 0$)} \label{Corr2}

In this section we discuss the effects of the $\al'^2$-terms localized on the 8 D7-branes and the O7-plane when $M\neq 0$. For this case one expects that the $\al'^2$-terms will give corrections to the supergravity background. Our goal is to investigate what the interpretation of these corrections are in the dual field theory. Our approach is similar to the analysis in \cite{Frolov:2001}, but there are some differences related to the lower-dimensional nature of the D7-branes.

\subsection{The supergravity background}
In this section we briefly review the type IIB supergravity solution obtained in \cite{Klebanov:2000a} (with some normalization factors taken from \cite{Herzog:2001}). This solution is an approximation to the solution obtained in 
\cite{Klebanov:2000b}, but it will be sufficient for our later purposes. As we will discuss below the solution is also a solution to the orientifolded theory.

The metric is given by  (to facilitate our later discussion we have written the solution using the variables introduced in \cite{Frolov:2001}, see \cite{Frolov:2001} for further details )
\be \label{met}
\D s_{10}^2 = \rL^2 [ \frac{e^{2t}}{\sqrt{t}}\D x_n \D x^n + \sqrt{t}(\D t^2 +  \D s_{T^{11}}^2)]\,,
\ee
where $n=0,\ldots,3$ and
\be \label{rL}
\rL^2 = \frac{9 g_{\mathrm{S}} (2M)\al'}{2\sqrt{2}}\,.
\ee
The solution also contains the following non-zero field strengths 
\bea \label{Fs}
&F_3 = M\al'\om_3 \,, & \qquad  B_2 = 3 g_{\mathrm{S}} M \al'(t-\tilde{t}) \om_2\,, \non \\ 
&F_5 = \D C_4 + B_2 \we F_3 = \cF_5 + *\cF_5\,, & \qquad  \cF_5 = 27 \pi \al'^2\bar{N}_{\mathrm{eff}}(t)\mathrm{vol}(T^{11})\,,
\eea
where $\mathrm{vol}(T^{11})$ is the volume form on $T^{11}$ and 
\be
\tilde{t} = \frac{2\pi (2N)}{3g_{\mathrm{S}}(2M)^2} + {\ts \frac{1}{4}} \,, \qquad \bar{N}_{\mathrm{eff}}(t) = \frac{3 g_{\mathrm{S}} (2M)^2 }{2\pi}\left(t-\frac{1}{4}\right)\,,
\ee
and  $\om_2$ and $\om_3$ are given by
\bea
\om_2 &=& \sin \tha_1 \D \tha_1 \we \D \phi_1 - \sin \tha_2 \D \tha_2 \we \D \phi_2 \,,\non \\
\om_3 &=& (\D \psi + \cos\tha_2 \D\phi_1+ \cos\tha_2 \D \phi_2)\we \om_2\,.
\eea
In writing the above solution we have anticipated the fact that the number of branes on the cover space of the $\Sp(2N{+}2M){\times}\Sp(2N)$ theory is twice as large compared to the $\SU(N{+}M){\times}\SU(N)$ model. The fluxes and the number of physical branes are however the same since the volumes of the cycles over which one integrates are half as large, e.g. $\frac{1}{4\pi^2\al'}\int_{S^3/\mathsf{Z}\kern -3pt \mathsf{Z}_2} F_3 = M$.

As was discussed in section \ref{Sori} the spacetime part of the orientifold operation interchanges $(\phi_1,\tha_1)$ and $(\phi_2,\tha_2)$ which implies $G_{MN}\rar G_{MN}$, $\om_2 \rar -\om_2$ and $\om_3 \rar -\om_3$. In the solution (\ref{Fs}) one has $B_2 \propto \om_2$, $F_3 \propto \om_3$ and $\cF_5\propto \om_2 \wedge \om_3$. In addition to the spacetime part the orientifold operation also involves the action of $\Om(-1)^{F_L}$. The total orientifold operation is such that it leaves the form fields in (\ref{Fs}) invariant.  Thus, as was also discussed in \cite{Imai:2001,Naculich:2002}, the above solution is a solution in the orientifolded theory since it is invariant under the orientifold action.

\subsection{The beta function and the RG flow}
We now discuss the beta function and the associated renormalization group flow in the $\Sp(2N{+}2M){\times}\Sp(2N)$ field theory and how these concepts appear in the dual supergravity solution discussed above.

\medskip
{\it Field theory}
\medskip

The field theory beta-function relevant to our discussion is the Shifman-Vainshtein \cite{Shifman:1986} beta function (see e.g.~\cite{Herzog:2001,Frolov:2001} for an explanation of this fact\footnote{Recently it has been shown \cite{Imeroni:2002} that it is possible to obtain the complete NSVZ beta function \cite{Novikov:1983} from the supergravty solution (see also \cite{DiVecchia:2002} and references therein).}):
\be
\beta_g = 
\frac{\D}{\D \log(\La/\mu) } \left( \frac{4\pi^2}{g^2} \right) = \frac{1}{4} \Big[ 3 T(\mathrm{adj}) - \sum_i  T(R_i)(1-2\ga_i) \Big]  \,,
\ee
where the sum runs over the matter superfields and $T(R_i)$ is the index of the representation $R_i$ of the gauge group. For $\Sp(2N)$, $T(\mathrm{adj})=2N+2$ and $T(\Yfund)=1$. Using these results together with the matter content given in section \ref{Sori}, the beta functions for the gauge couplings (labeled $g_1$ and $g_2$) of the two factors of the $\Sp(2N+2M){\times}\Sp(2N)$ gauge group can be conveniently written as 
\bea \label{betas}
\frac{\D}{\D \log(\La/\mu) } \left( \frac{4\pi^2}{g_1^2}+\frac{4\pi^2}{g_2^2} \right) &=& 2\bN \De + 2 \de \,, \non \\
\frac{\D}{\D \log(\La/\mu) } \left( \frac{4\pi^2}{g_1^2}-\frac{4\pi^2}{g_2^2} \right)&=&  M (3-\De) \,,
\eea
where $\bN = N + M/2$ (note that $\bar{N}$ is simply the effective number of physical D3-branes in the defining orientifolded conifold background) and $\de$ and $\De$ are defined via $\ga_B = -\frac{1}{4} + \frac{\De}{2}$ and $\ga_F = -\frac{1}{4}+ \frac{\de}{2}$, where $\ga_B$ is the anomalous dimension of the bifundamental matter multiplets and $\ga_F$ is the anomalous dimension of the fundamental matter multiplets. When $M=0$, $\De$ and $\de$ are both zero (i.e.  $\ga_B = -\frac{1}{4}=\ga_F$) and both beta functions vanish. When $M$ is not zero $\De$ and $\de$ can be expanded in even powers of $\frac{M}{\bN}$, which is assumed to be small (the argument why only even powers appear is completely analogous to the one given in \cite{Frolov:2001}). To leading order in $M$ one finds that only the difference of the gauge couplings flows: $\frac{\D}{\D \log(\La/\mu) } \left( \frac{4\pi^2}{g_1^2}-\frac{4\pi^2}{g_2^2} \right) =  3M$.

For completeness we note that for the $\Sp(2N{+}2M){\times}\SO(2N{+}2)$ theory discussed in section \ref{Sori} one finds (using that for $\SO(N)$, $T(\mathrm{adj})=N-2$ and $T(\Yfund)=1$)
\bea
\frac{\D}{\D \log(\La/\mu) } \left( \frac{4\pi^2}{g_1^2} + \frac{4\pi^2}{g_2^2} \right) &=& 2\bN \De  \,,\non \\
\frac{\D}{\D \log(\La/\mu) } \left( \frac{4\pi^2}{g_1^2} - \frac{4\pi^2}{g_2^2} \right)&=&  3M-(M+1)\De\,,
\eea
where $\bar{N}=N + (M{+}1)/2$. Note that for this model there is no longer a strict symmetry between the two factors of the gauge group.

\medskip
{\it String theory}
\medskip

The field theory gauge couplings are encoded in the gravity dual via the relations \cite{Klebanov:1998,Morrison:1998} 
\bea \label{sb} 
(\frac{4\pi^2}{g_1^2} + \frac{4\pi^2}{g_2^2}) &=& \frac{\pi}{g_{\mathrm{S}}}e^{-\Phi} \,,\non \\
(\frac{4\pi^2}{g_1^2} - \frac{4\pi^2}{g_2^2}) &=& \frac{1}{g_{\mathrm{S}}}e^{-\Phi}[\frac{1}{2\pi\al'}\int_{S^2/\mathsf{Z}\kern -3pt \mathsf{Z}_2} B_2 - \pi]\,.
\eea
In the background (\ref{met}), (\ref{Fs}) one finds \cite{Herzog:2001} (using $\int_{S^2/\mathsf{Z}\kern -3pt \mathsf{Z}_2}\om_2 = 2\pi$) that to leading order  in $\frac{M}{\bar{N}}$, $\frac{\D}{\D t }(\frac{4\pi^2}{g_1^2} - \frac{4\pi^2}{g_2^2})= 3M$; hence by comparison with the above field theory result one deduces that $t$ is to be identified with $\log(\La/\mu)$ (up to an additive constant). Hence, there is a complete matching between the beta function obtained from field theory with that obtained from string theory \cite{Herzog:2001}; on the string theory side the parameter controlling the RG flow is identified with the radial variable $t$ appearing in the metric (\ref{met}).

In \cite{Frolov:2001} a consistent solution to the ten-dimensional field equations in the $\mathit{adS}_5{\times}T^{11}$ background in the presence of the $\al'^3R^4$ corrections was obtained by taking the deformations of the fields to depend only on the radial coordinate $t$. Furthermore, under the assumption that the relations (\ref{sb}) still hold for the case when $\Phi$ and $B_2$ no longer are constant, it was shown that the corrections could be interpreted in the dual field theory as corrections to the anomalous dimensions of the matter fields.  A natural guess based on the results in \cite{Frolov:2001} is that the $\al'^2$ corrections localized on the D7-branes and the O7-plane should also have a dual interpretation in terms of corrections to the anomalous dimensions of the matter fields in the field theory. 
However, in trying to make this more precise one encounters a puzzle.  An ansatz where the deformation only depends on the radial variable, $t$, is not possible since the presence of the D7-branes and the O7-plane will lead to $\de$-function sources in the equations of motion, as was discussed in section \ref{Corr1}. These sources explicitly depend on the angular coordinates of $T^{11}$; therefore the solution will also have to depend on the angles of $T^{11}$. On the other hand, from the field theory side one expects that $\Phi$ and $\int B_2$ should depend only on one variable which describes the RG flow. 

The resolution we would like to propose is that it is the zero modes of the KK expansion in modes on $T^{11}$ that should appear on the right hand side of the equations (\ref{sb}). This is a natural choice since such modes are always present independently of the choice of the compact manifold $X_5$ in the product $\mathit{adS}_5{\times}X_5$.  Furthermore, they are the only modes that do not depend on the angles on $T^{11}$.

\subsection{The corrections to the equations of motion}
As mentioned above, because of the presence of the $\de$-function source, the deformation of the metric induced by the presence of the $\al'^2$ terms will necessarily have to depend on the angles of $T^{11}$ in addition to the dependence on $t$. Thus, to get the complete solution one would have to include all modes on $T^{11}$ in the ansatz. However, since we are only interested in the zero-modes for $\Phi$ and $\int B_2$ it is natural to expect that the dominant contribution can be obtained by considering only the lowest lying modes in the mode expansion of the metric. The modes we will include in the ansatz for the metric are the same ones as in \cite{Frolov:2001}. Since the functions $w$, $y$ and $z$ in this ansatz only depend on $t$, including only these modes is equivalent to dimensionally reducing the system to an effective five-dimensional theory. With this in mind we choose our ansatz for the deformed metric to be of the same form as in \cite{Frolov:2001}:
\bea
\D s_{10}^2=\rL^2\Big[ \frac{e^{2t}}{\sqrt{t}}\D x_n \D x^{n} + \sqrt{t}\,e^{-2 z}\bigg\{e^{10 y }\D t^2 + \frac{e^{2y-8w}}{9}(\D \psi + \cos\tha_1 \D \phi_1 + \cos \tha_2 \D\phi_2)^2  \non \\ +  \frac{e^{2y +2w}}{6}(\D\tha_1^2 + \sin^2\tha_1\D\phi_1^2 + \D\tha_2^2 + \sin^2\tha_2\D\phi_2^2)\bigg\}\Big] \,,
\eea
where, $y$, $w$, and $z$ are of order $\al'^2$. Similarly, we will use the following ansatz for the NSNS two-form: $B_2 = 3 g_{\mathrm{S}} M\al'(t- b) \om_2 $, where the deformation $b$ again depends only on $t$.

The $R^2$ part of the action for the combined system of 8 D7-branes and one O7-plane becomes (after rescaling the background metric as $G \rar \rL^2 G$ and $C_4$ as $C_4 \rar 4\rL^4 g_{\mathrm{S}}^{-1}C_4$, where  $\rL^4$ was defined in (\ref{rL}))
\bea  \label{R2tot2}
S_{\mathrm{tot}} &=&  \mu_7 \rL^4 (\pi\al')^2 \Bigg\{ \int \sqrt{-g}e^{-\Phi}\Big[ R_{\rho\la\mu\nu}R^{\rho\la\mu\nu} - 2 R_{\mu\nu}R^{\mu\nu}  -\; R_{ij\mu\nu}R^{ij\mu\nu} + 2R_{ij}R^{ij}\Big] \non \\ &&\qquad \quad +\; 2 g_{S}^{-1}  \int C_4 \we [  \mathsf{R}^{\rho}{}_{\la} \we \mathsf{R}^{\la}{}_{\rho} - \mathsf{R}^i{}_j \we \mathsf{R}^j{}_i ] \Bigg\} \,.
\eea
As in section \ref{Corr1} our analysis will be restricted to these terms since the $F_5$ dependence is not known. Since $H_3$ is also non-zero in the background (\ref{met}), (\ref{Fs}) we would like to know the dependence of the effective action on this field. Such terms have not been determined in the literature (see, however, \cite{Scrucca:2000} for a discussion of such terms in the Wess-Zumino part of the D-brane action; note that the conditions on $H_3$ used in that paper are satisfied for  $H_3$ in (\ref{Fs})). Based on the bulk analysis one might have expected that there was a simple rule of the form $R \rar R + D H_3$ from which the $DH_3$ terms could be obtained from the $R^2$ terms. As briefly discussed in the appendix this does not appear to be the case. Because of our lack of knowledge about the $H_3$ terms, all terms involving $H_3$ will be neglected below.

The equations of motion, with the above simplifying assumption, become (using the expressions given in \cite{Frolov:2001})
\bea
 8\frac{e^{4t}}{t} \Big[ \frac{1}{2} b'' + 2b'-\frac{2}{t}b' -\frac{2}{t}b-\frac{1}{2} \phi' +\frac{1}{2t}\phi-2\phi && \non \\  \qquad-2w'-8w+\frac{2}{t}w  -2y'-8y+\frac{2}{t}y+2z'-8z+\frac{2}{t}z\Big] &=& -\frac{2\ka^2}{\rL^8 } \frac{\de S_{\mathrm{brane}}}{\de b} \,,\non \\ 
 e^{4t} \Big[ \phi'' + 4\phi' -\frac{4}{t}\phi +\frac{4}{t}b' -\frac{16}{t}y -\frac{16}{t}w \Big] &=& -\frac{2\ka^2}{\rL^8 }\frac{\de S_{\mathrm{brane}}}{\de \phi} \,,\non \\ 
 8e^{4t} \Big[ -5y'' -20 y'+160y -\frac{8}{t}y +\frac{2}{t}b'-\frac{2}{t} \phi -\frac{8}{t}w \Big] &=& -\frac{2\ka^2}{\rL^8 }\frac{\de S_{\mathrm{brane}}}{\de y} \,,\non \\ 
 16 e^{4t} \Big[ z'' +4 z'-32z +\frac{12}{t}z -\frac{2}{t^2}z-\frac{1}{t} b' -\frac{8}{t}b + \frac{2}{t^2}b \Big] &=& -\frac{2\ka^2}{\rL^8 }\frac{\de S_{\mathrm{brane}}}{\de z} \,,\non \\ 
 8 e^{4t} \Big[ 5w'' +20 w'-60w -\frac{8}{t}w +\frac{2}{t}b'-\frac{2}{t} \phi -\frac{8}{t}y \Big] &=& -\frac{2\ka^2}{\rL^8 }\frac{\de S_{\mathrm{brane}}}{\de w}\,.
\eea 
Here the quantities appearing on the left hand depend only on $t$, and the quantities on the left hand side are assumed to have been integrated over the angles and consequently only depend on $t$.   
We have found that there is a field redefinition scheme were the modifications to the equations of motion for the metric have the following asymptotic form for large $t$: 
\be \label{ywzscal}
\frac{\de S_{\mathrm{brane}}}{\de z} \sim \mu_7 g_{\mathrm{S}}^{-1} \al'^2 \rL^4 e^{4t}t^{-2}\,, \quad \frac{\de S_{\mathrm{brane}}}{\de y} \sim  \mu_7 g_{\mathrm{S}}^{-1} \al'^2 \rL^4 e^{4t}t^{-2}\,, \quad \frac{\de S_{\mathrm{brane}}}{\de w} \sim \mu_7 g_{\mathrm{S}}^{-1} \al'^2 \rL^4 e^{4t} t^{-2}.
\ee
More precisely, we have verified that there is a field redefinition scheme for which this is true. Naively, one would find the scaling $t^{-1}$. Note that the fact that there is a scheme in which the $t^{-1}$ terms vanish implies that the corrections vanish when $M\rar 0$. The reason that there is a field redefinition scheme for which the corrections to the equations of motion for the metric vanishes in this case is that the deformation of the metric is of a restricted form, whereas in section \ref{Corr1} the deformation was general. For $b$ and $\phi$ the following asymptotic scaling
\be 
\frac{\de S_{\mathrm{brane}}}{\de \phi} \sim \mu_7 g_{\mathrm{S}}^{-1}  \al'^2 \rL^4 e^{4t} t^{-3}\,, \qquad \frac{\de S_{\mathrm{brane}}}{\de b} \sim \mu_7 g_{\mathrm{S}}^{-1}  \al'^2 \rL^4 e^{4t} t^{-3}\,,
\ee 
appears to be required for consistency. We have verified that there is a field redefinition scheme for which the first equation holds (in addition to the equations in (\ref{ywzscal})). Due to our lack of knowledge of the $H_3$ dependence the second relation is conjectural. 
The asymptotic solution consistent with the above scaling is
\be
z \sim  g_{\mathrm{S}}\frac{\al'^2}{\rL^4}t^{-2}\,, \quad y \sim  g_{\mathrm{S}}\frac{\al'^2}{\rL^4}t^{-2}\,, \quad w \sim  g_{\mathrm{S}}\frac{\al'^2}{\rL^4}t^{-2}\,, \quad \phi \sim   g_{\mathrm{S}}\frac{\al'^2}{\rL^4}t^{-2}\,, \quad b \sim   g_{\mathrm{S}}\frac{\al'^2}{\rL^4}t^{-1} \,,
\ee
where we have used the above asymptotic scalings together with $2\ka^2=(2\pi)^7g_{\rm S}^2\al'^4$ and $\mu_7 = [2(2\pi)^7\al'^4]^{-1}$. As in \cite{Frolov:2001} we find that the asymptotic scaling is the minimal choice consistent with the absence of corrections when $M\rar0$.

By comparison with (\ref{betas}), (\ref{sb}) we find from the first relation  that to leading order
\be
\De + \frac{1}{\bN}\de = -\frac{2\pi^2}{\la}\phi'\,.
\ee
Evaluating the derivative at the point $t_{\mathrm{eff}}$ for which $N_{\mathrm{eff}}(t_{\mathrm{eff}}) = \bar{N}$ (see \cite{Frolov:2001} for more details) and using the scalings $\frac{\al'}{\rL^2} \sim \frac{1}{\la}\frac{\bN}{M}$ and $\frac{1}{t_{\mathrm{eff}}} \sim \la \left(\frac{M}{\bN}\right)^2$ together with the above asymptotic form of $\phi$ we find 
\be
\De + \frac{1}{\bN}\de \sim g_{\mathrm{S}} \left(\frac{M}{\bN}\right)^4\,.
\ee
Since this relation depends on both $\De$ and $\de$ we are not able to determine them separately without further input. It does seem natural to assume that only $\de$ (and not $\De$) should be affected by the corrections on the D7's in which case one finds  $ \de \sim \la \left(\frac{M}{\bN}\right)^4$. 
At first sight this dependence on the 't Hooft coupling looks strange, but it is consistent with the fact that the expansion is in terms of $\frac{1}{t} \sim \la(\frac{M}{\bar{N}})^2$ and $\frac{\al'}{\rL^2} \sim \frac{1}{\la} \frac{\bar{N}}{M}$ (which imply that $\frac{1}{\la}$ and $\frac{M}{\bar{N}}$ are also small).  For consistency of the interpretation that the $\al'^2$-corrections only affect $\de$ with the second relation coming from (\ref{betas}), (\ref{sb}) one needs to assume that $b' = \phi$, a relation which was needed in the bulk analysis for consistency. However, if there is a correction to both $\de$ and $\De$ at order $\al'^2$ then this restriction may not be needed. 

Let us also mention that even for the bulk $\al'^3R^4$ terms there is a slight difference compared to \cite{Frolov:2001} since now not only $\De$ appears in the beta function but also $\de$. However, if our above suspicion that the corrections to $\de$ arise form the corrections on the brane is correct, we do not expect $\de$ to receive corrections from the bulk $\al'^3$ terms, but only from the $\al'^3$ terms localized on the brane.

\section{Discussion}
In this paper we have studied the $\al'^2$ terms localized on the O7-plane and eight D7-branes in a particular orientifold of $\mathit{adS}_5{\times}T^{11}$. This orientifold is dual to an $\cN{=}1$ $\Sp(2N{+}2M){\times}\Sp(2N)$ gauge theory. 

In section \ref{Corr1} we argued that when $M=0$ the $\al'^2$ terms should give no corrections (modulo field redefinitions) to the supergravity equations of motion. 
Evidence for this expectation comes from the adS/CFT correspondence. 
In particular, the paramaters $g_{\mathrm{S}}$ and $N$ which are related to the dilaton and the radius of $\mathit{adS}_5{\times}T^{11}$ have to remain free parameters in the full string theory (see e.g.~\cite{Frolov:2001}). Furthermore, the symmetries of the gauge theory (which are related to the isometries of $\mathit{adS}_5{\times}T^{11}/\Z_2$) have to be preserved. 
Taken together these facts indicate that all corrections to the supergravity background are absent. However, to our knowledge it has not been rigorously proven that the adS/CFT correspondence implies that all corrections to the supergravity background have to be absent, although it seems very plausible that that is the case. (If there were residual corrections not removable by field redefinitions, it would pose a challenge to interpret them in the context of the dual conformal field theory.)  
Due to the lack of knowledge about the $F_5$-dependent terms we were unable to completely prove that there are no corrections at order $\al'^2$, although our results are quite suggestive (see section \ref{Corr1} for further details). 
One can also ask the reverse question, i.e.~requiring the $\al'^2$ terms to give rise to no corrections for the orientifold background, what are the possible $R^2$ terms?  As we have discussed the number of equations that have to be satisfied in order for there to be no corrections is fairly large so one would expect strong restrictions. The field redefinition ambiguity limits the usefulness of such an approach and one would also have to make an ansatz for the unknown $F_5$ terms. If one were to find an analogue of the prescription of replacing the bulk Riemann tensor with the Weyl tensor for the terms on the brane the restrictions would be even stronger. One stumbling block towards realizing this program is the fact that some of the possible $R^2$ terms have vanishing first order variations in the orientifold background. However, it seems that none of the terms with this property actually appear in the action, so there may be some symmetry which forbids them. The conclusion we can draw from the above discussion is that the restrictions are quite strong, but that one needs additional input from other approaches to be able to use them constructively to constrain e.g.~the $R^3$ terms at the next order. 

In section \ref{Corr2} we studied the $\al'^2$ corrections in the cascading case ($M\neq 0$). Using  methods similar to the ones in \cite{Frolov:2001} we found that there are corrections to the background, which following \cite{Frolov:2001} we interpreted on the field theory side as corrections to the anomalous dimensions of the matter fields.

There is clearly room for additional work extending and refining our results. In particular, the fact that we do not know the complete expression for the D-brane action at order $\al'^2$ make some of our results speculative. For the bulk terms considered in \cite{Frolov:2001} the dependence on $F_5$ was also unknown; in our case the situation is similar, except that the $F_5$ dependent terms appear to play a more important role. The determination of the $F_5$ terms is expected to be very involved, but they are probably more accessible than the corresponding bulk terms (for instance, the $DF_5 DF_5$ terms in the DBI part of the action can be constrained by the string scattering amplitudes calculated in \cite{Gubser:1996}) and there is some hope that it will be possible to determine the complete expression for the effective action at order $\al'^2$ (including also the $H_3$ terms), which would allow one to unambiguously verify our results.

\section*{Acknowledgements}
We would like to thank S. Naculich for collaboration in the early stages 
of this project and for comments on the manuscript. 
For some of the calculations we found the GrTensor package \cite{GrTensor} to be useful. HJS would like to thank the string theory group and the Physics Department of Harvard University for their hospitality extended over a long period of time.

\appendix

\section{Some string scattering amplitudes}

\subsection{Scattering off a D-brane}
 One way to determine the amplitude for the scattering of bulk states off a D-brane is by using the boundary state formalism. The amplitude for the scattering of an NSNS state off a D-brane in this formalism is given by
\be \label{Dscat}
\mathcal{A} \propto {}^{(-1,-1)}\langle k; \xi|V^{(0,0)}_{q;\ze}\De |B\rangle \,,
\ee
where $|B\rangle$ is the usual boundary state \cite{Callan:1987a,Callan:1987b,Callan:1988}, ${}^{(-1,-1)}\langle k, \xi|$ is the NSNS state in the $(-1,-1)$ picture with polarization tensor $\xi_{MN}$ and momentum $k^M$, $\De$ is the string propagator $\int_0^1\D y \,y^{L_0+\tilde{L}_0-1}$ (throughout we suppress all ghosts), and $V^{(0,0)}_{q;\zeta}$ is the vertex operator in the (0,0) picture for a NSNS state with polarization tensor $\ze_{MN}$ and momentum $q^M$, i.e. 
\be
V^{(0,0)}_{\zeta;q} = \ze_{MN}:(\pa X^{M} + q_{P}\Psi^{P}\Psi^{M})(\bar{\pa} \tilde{X}^{N} + q_{S}\tilde{\Psi}^{S}\tilde{\Psi}^{N})e^{iq_{S}X^{S}}: \,.
\ee 
In our conventions, momentum conservation in the longitudinal directions implies that  $k^M + D^M{}_N k^N = q^M + D^M{}_N q^N$, where $D_{M}{}^{N}$ is a diagonal matrix whose entries are $+1$ in the world-volume directions and $-1$ in the transverse directions. The scattering amplitude (\ref{Dscat}) is  \cite{Garousi:1996}\footnote{This expression was originally derived \cite{Klebanov:1996,Garousi:1996,Hashimoto:1997} using a slightly different approach from the one discussed here. As pointed out in \cite{Corley:2001} there are minor errors in the expression in \cite{Garousi:1996}; our expression confirms the formula in \cite{Corley:2001}. }
\be \label{AD}
\mathcal{A} \propto \frac{\Ga(2s)\Ga(-t/2)}{\Ga(2s-t/2+1)}[2s a_1 + {\ts \frac{t}{2}}a_2] \,,
\ee
where
\bea \label{aD}
a_1 &=& {\ts \frac{\alpha'}{2}} [(D\zeta)(q\xi q) + (D\xi)(k\zeta k)] + 2 s (\zeta \xi^T) + {\ts \frac{\alpha'}{2}}[k\zeta D\xi q + q\xi D \zeta k] \non \\
&&  - {\ts \frac{\alpha'}{2}}[ q \xi \zeta^T k + k \zeta^T \xi q +  q D \xi \zeta^T k + k D \zeta \xi^T q + k \zeta^T \xi D q + q \xi^T \zeta D k ]\,, \non \\
a_2 &=& {\ts \frac{\alpha'}{2}}[ (D \zeta)(q D \xi D q - q\xi q) + (D \xi)(k D \zeta D k - k \ze k) - 2 s (\ze \xi^T) \non \\ 
&& - {\ts \frac{\alpha'}{2}}[ q D \ze D \xi D k + k D \xi D \ze D q] + 2s(D \ze D \xi) - (2s{-}t/2)(D\ze)(D\xi) \non \\
&& + {\ts \frac{\alpha'}{2}}[kD \xi \ze^T D q + k D \xi^T \ze D q] \,.
\eea
We have defined $t = \al' k q$ and $s = \frac{\al'}{4} k D k = \frac{\al'}{4} q D q$ and use a notation were the index contractions are suppressed, e.g.~$k D \ze \xi^T q = k_M D^{MN}\xi_{PN}q^P$. 
The amplitude (\ref{AD}), (\ref{aD}) is invariant under the interchange of the two states, $k \leftrightarrow q$; $\xi \leftrightarrow \ze$, and is also invariant under the gauge transformation $\xi_{MN} \rar \xi_{MN} + k_M \xi_N + k_N \xi_M$ (when $\xi_{MN}$ is symmetric).
The above scattering amplitude was used in \cite{Bachas:1999} (see also \cite{Fotopoulos:2001}) to determine the $\al'^2 R^2$ terms in the DBI part of the D-brane effective action.

The above amplitude can be used to restrict the form of the $R D^2 \phi$ terms in the DBI part of the D-brane effective action.  The dilaton vertex operator is 
\be
\xi_{MN}(k) \propto \phi(k)(\eta_{NM} - \ell_{M}k_N - \ell_N k_M)\,,
\ee
where $\ell_M k^M =1$. From (\ref{AD}), (\ref{aD}) one finds that the $\al'^2$ part of the amplitude for one dilaton (with polarization $\xi_{MN}$ and momentum $k_M$) and one graviton (with polarization tensor $\ze_{MN}$ and momentum $q_M$) is
\be
\mathcal{A} \propto 2  (p-3) \phi(k) \Big\{(2s-t/2)[(k\ze k) -(t/2)(D\ze)] + (t/2)(k D\ze D k)\Big\}\,.
\ee 
The following field-theory terms 
\be \label{PhiR}
S \propto \int \sqrt{-g} [ a_1  L^{M_1M_2}L^{N_1N_2} + a_2 N^{M_1M_2}N^{N_1N_2} + bL^{N_1M_1}L^{M_2N_2}] N^{PQ}R_{PM_2 Q N_2} D_{M_1}\pa_{N_1}\Phi\,,
\ee
with $a_2 = p-3$ and $a_1 + 2b=-a_2$ reproduce the string scattering result. We should remember that the comparison is done in a particular frame; however, the terms induced by Weyl rescaling the metric in the action (\ref{R2tot}) are of the same form as above. More precisely, via the Weyl rescaling one gets terms with $a_1 = (p-3)\ga$, $b = 2\ga$ and $a_2 = -(p-1)\ga$, where $\ga$ is a normalization constant. Note that the scattering amplitude makes no distinction between $L^{N K}D_{N}R_{M P}\pa_{L}\Phi$ and $L^{N K} R_{M P}D_{N}\pa_{L}\Phi$. To the above result one can also add terms proportional to $D^M\pa_M\phi$ or $R_{MN}=G^{PQ}R_{PMQN}$ since such terms are removable by field redefinitions. For our purposes the ambiguities in (\ref{PhiR}) will not be important since for the special background we consider there is a choice of field redefinition scheme in which the terms in (\ref{PhiR}) give vanishing contributions to the equation of motion for $\Phi$ no matter what the values of $a_1$, $a_2$ and $b$ are. 

The above amplitude (\ref{AD}), (\ref{aD}) can also be used to obtain restrictions on the possible $D H_3 D H_3$ terms in the DBI part of the D-brane effective action. We have found that the string scattering amplitude (\ref{AD}), (\ref{aD}) is consistent with the following terms in the effective action 
\bea  \label{DHDH}
S = \mu_{p}{\ts \frac{\pi^2}{6} \frac{\al'^2}{8} } \int \D^{p+1} x \sqrt{-g}\Big[ L^{N_1 Q_2}L^{N_2Q_1}L^{M_1 M_2}N^{P_1 P_2}  - \half N^{N_1 Q_2}L^{N_2Q_1}L^{M_1 M_2}L^{P_1 P_2} \non \\ - \, \half L^{N_1 Q_2}N^{N_2Q_1}N^{M_1 M_2}N^{P_1 P_2} \Big]D_{N_1}H_{M_1P_1Q_1}D_{N_2}H_{M_2P_2Q_2}\,,
\eea
where $L^{MN}$ and $N^{MN}$ are the same as before, cf.~(\ref{LN}). Terms proportional to the lowest order equation of motion for $H_3$: $0 = G^{MN}D_N H_{M PQ} = (L^{MN} + N^{NM})D_N H_{M PQ}$ are removable by field redefinitions. It is important that apart from the field redefinition ambiguity, there are several additional ambiguities in (\ref{DHDH}) related to the fact that the scattering amplitude can not distinguish between e.g.~$L^{K N_1} D_{N_1}H_{M_1P_1Q_1}D_{N_2}H_{M_2P_2Q_2}$ and  $-L^{K N_1} H_{M_1P_1Q_1}D_{N_1}D_{N_2}H_{M_2P_2Q_2}$ since an integration by parts relates the two (modulo terms which do not contribute to the scattering amplitude we considered). Apart from the ambiguities related to integrations by parts it is also not possible to determine terms of the form $H_3^2 R$ and $H_3^4$ from the scattering amplitude (\ref{AD}). The determination of such terms would require the calculation of amplitudes with additional vertex operator insertions. One thing to note about the above result is that there does not appear to be a simple way to combine the $DH_3DH_3$ terms with the $R^2$ terms via a replacement rule of the form $R \rar R + D H_3$ (except for the D9-brane for which the $DH_3DH_3$ terms are removable by field redefinitions).

\subsection{Scattering off an O-plane}

Above we discussed the scattering of an NSNS state off a D-brane. The scattering of an NSNS state off an O$p$-plane is obtained in an analogous way  by simply replacing the boundary state in the above amplitude with the cross-cap state, $|C\rangle$, i.e.
\be \label{Oscat}
\mathcal{A} \propto {}^{(-1,-1)}\langle k, \xi|V^{(0,0)}_{q;\zeta}\De |C\rangle\,.
\ee
The cross-cap state associated with an O$p$-plane satisfies \cite{Callan:1987b}
\be
\al^{M}_n|C\rangle = -(-1)^n D^M{}_N\tilde{\al}^{N}_{-n} |C\rangle\,, \qquad
b^{M}_{r}|C\rangle = i \eta (-1)^r D^M{}_N\tilde{b}^{N}_{-r}|C\rangle\,,
\ee
where $\eta$ is the spin structure and $\al_n^N$ and $b^M_r$ are the usual oscillator modes.
The amplitude (\ref{Oscat}) becomes
\be \label{AO}
\mathcal{A} \propto \Big[ a_1 I_1 + a_2 I_2 + a_3 I_3 \Big]\,,
\ee
where
\bea \label{aO}
a_1 &=& {\ts \frac{\alpha'}{2}}[(\xi D)(k\ze k) + (\ze D)(q\xi q)] + 2s (\xi \ze^T) + {\ts \frac{\alpha'}{2}}[k \ze D \xi q + q \xi D \ze k] \non \\
&& - {\ts \frac{\alpha'}{2}}[ k\ze \xi^T q + k \ze^T \xi q + k \ze \xi^T D q + k \ze^T \xi D q + k D \ze \xi^T q + k D \ze^T \xi q ] \,,\non \\
a_2 &=& {\ts \frac{\alpha'}{2}}[ (\xi D)(k D \ze D k) + (\ze D) (q D \xi D q)] + 2 s (\ze D \xi D) + {\ts \frac{\alpha'}{2}}[k D \ze \xi^T D q + k D \ze^T \xi D q] \non \\ 
&&+ {\ts \frac{\alpha'}{2}}[k \ze D \xi q + q \xi D \ze k - q D \ze D \xi D k - k D \xi D \ze D q] \,, \non \\
a_3 &=& -(2s-1)(\xi D)(\ze D) \,,
\eea
and
\bea \label{Is}
I_1 &=& \int_0^1 \D y \,y^{-t/2-1}(1+y)^{2s-1} \,, \non \\
I_2 &=& \int_0^1 \D y \,y^{-t/2}(1+y)^{2s-1} \,, \non \\
I_3 &=& \int_0^1 \D y \,y^{-t/2}(1+y)^{2s-2} \,.
\eea
The asymptotic expansion of the $I_i$'s for small momenta obtained by integrations by parts contains terms proportional to $\ln 2$. Such terms are not expected to appear in the amplitude. The reason for the appearance of these terms is that the above amplitude is not gauge invariant unless the following relations hold: $(t/2)I_1 = (2s -t/2)I_2$ and $(2s-1)I_3 = (t/2)I_1$. By integrations by parts in the above expressions (\ref{Is}) it can be shown that these relations only hold up to boundary terms. 
The reason why the amplitude (\ref{AO}), (\ref{aO}) is not gauge invariant  can be traced to the fact that it was derived using the standard vertex operators in (\ref{Oscat}), however, on the cross-cap these are only gauge invariant  up to boundary terms. For discussions of boundary (or contact) terms in string scattering amplitudes and their relation to gauge invariance, see e.g.~\cite{Gutperle:1997}. 
Rather than rederiving the amplitude using the correct gauge-invariant vertex  we will try to identify the required boundary terms by demanding that the amplitude be gauge invariant and that it correctly reproduces the leading order terms. 
We will also demand that the $\al'\ln2$ terms should be absent since their presence would be in conflict with the fact that such terms are not present in the effective action for type I string theory (type I string theory is essentially 1 O9-plane and 32 D9-branes and since the D9-brane effective action does not contains $\al'\ln 2$ terms, such terms can not be present on the O9-plane either).  
We have found that if $(2/t)(2s -t/2)I_2 = (2/t)(2s-1)I_3 = I_1 = \int_0^1 \D y \,y^{-t/2-1}(1+y)^{2s-1} + \left[(1+y)^{2s-1}\right]^1_0 \sim -\frac{2}{t} -(t/2) \frac{\pi^2}{12} + \ldots$ all requirements are satisfied. For instance, a check of the amplitude is that it correctly reproduces the lowest order terms coming from the bulk supergravity action and from the terms localized on the O-plane. The diagrams that contribute to the two-graviton scattering are

\bigskip

\begin{figure}[h]
\begin{center}
\begin{tabular}{c@{\extracolsep{50pt}}cc} \includegraphics[height=2cm]{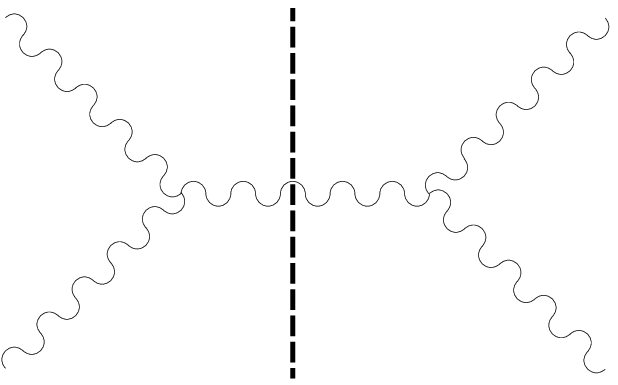}  &  \includegraphics[height=2cm]{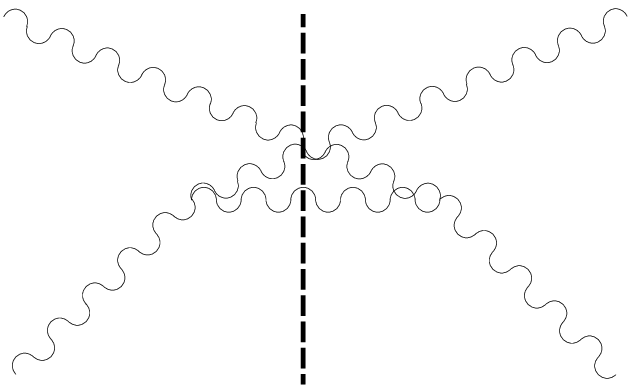}  &  \includegraphics[height=2cm]{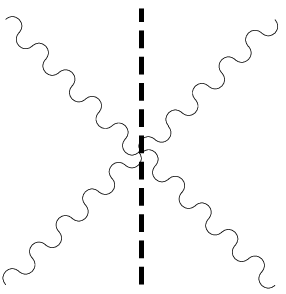} \end{tabular} \\
\flushleft{\small {\bf Figure:} Diagrams contributing to the scattering of a gravitons off an  orientifold plane. For clarity the diagrams are drawn on the cover space, i.e.~the mirror process has been included.}
\end{center}
\end{figure}

Calculating these diagrams with the proper symmetry factors gives (the third diagram comes from the DBI part of the terms localized on the orientifold plane)
\bea \label{Ais}
\mathcal{A}_1&\propto& \frac{2}{t}\Big[ 2(s-t/2)(\xi\ze) - t(\xi D\ze) + 2q\xi D\ze k + (\xi D)(k\ze k) + (\ze D) (q\xi q) \non \\ && \quad + 2q\xi \ze k  -2k\ze \xi D k - 2q\xi \ze D q\Big] \,,\non \\
\mathcal{A}_2&\propto& \frac{1}{(2s-t/2)}\Big[2(t/2 -s)(D\xi D \ze) - 2(2s -t/2)(\xi D \ze) + (\xi D)(kD\ze Dk)  \non \\ &&  \qquad + (\ze D) (qD\xi Dq) + 2 q D \xi \ze D k + 2q D\xi D \ze D k  -2k D\ze D\xi D k - 2qD\xi D \ze D q \Big] \,, \non \\ 
\mathcal{A}_3 &\propto& \Big[2 (\xi D \ze D) + 2 (\xi \ze) + 4(\xi D \ze) - (\xi D)(\ze D)\Big]\,.
\eea 
Notice that the second amplitude is obtained from the first one via $\xi \rar D \xi D$; $k \rar D k$ and that the first amplitude is the same as the one calculated in \cite{Corley:2001}. The sum of the above amplitudes (\ref{Ais}) is gauge invariant under $\xi_{MN} \rar \xi_{MN} + k_M\xi_N + k_N \xi_M$. 
The string scattering amplitude is consistent with this result if $I_1 \sim \frac{2}{t}$, $I_2 \sim \frac{1}{(2s-t/2)}$ and $(2s -1)I_3 \sim 1$. 

{}From the above amplitude one can derive the $\al'^2R^2$ terms localized on the O$p$-plane. The result is given in (\ref{ODBI}) (the result is independent of $p$). Although the above discussion of the boundary terms was incomplete, it is important to note that a more proper treatment of the boundary terms will not affect the $\pi^2\al'^2$ terms (which are gauge invariant) and hence will not affect (\ref{ODBI}), since it is not possible to get factors of $\pi$ from boundary terms.

\begingroup\raggedright\endgroup

\end{document}